\documentclass[%
 reprint,
twocolumn, 
 amsmath,amssymb,
]{revtex4}

\usepackage{graphicx}% Include figure files
\usepackage{dcolumn}% Align table columns on decimal point
\usepackage{bm}% bold math
\usepackage{dsfont}
\usepackage{cases}
\usepackage{color}
\usepackage{comment}
\usepackage{blindtext}
\usepackage{float}
\usepackage{lipsum}
\usepackage{braket}

\begin{document}

\title{Quasi-particle residue and charge of the one-dimensional Fermi polaron}

\author{Giuliano Orso$^1$}
\email{giuliano.orso@u-paris.fr}
\author{Lovro Anto Bari\v{s}i\'c$^2$, Ekaterina Gradova$^2$, Fr\'ed\'eric Chevy$^{2,3}$}
\author{Kris Van~Houcke$^2$}
\email{kris.van.houcke@phys.ens.fr}

\affiliation{
    $^1$ Universit\'e Paris Cit\'e, Laboratoire Mat\'eriaux et Ph\'enom\`enes Quantiques (MPQ), CNRS, F-75013, Paris, France \\
    $^2$ Laboratoire de Physique de l'Ecole Normale Sup\'erieure, ENS, Universit\'e PSL,CNRS, Sorbonne Universit\'e, Universit\'e Paris Cit\'e, F-75005 Paris, France \\
    $^3$ Institut Universitaire de France (IUF), 75005 Paris, France
}

\date{\today}

\begin{abstract}
We consider a mobile  impurity coupled to an ideal Fermi gas in one spatial dimension through an attractive contact interaction. We calculate the quasi-particle residue $Z$ exactly, based on Bethe Ansatz and diagrammatic Monte Carlo methods, and with
varational Ansatz up to one particle-hole excitation of the Fermi sea. 
We find that the exact quasi-particle residue vanishes  in the thermodynamic limit as a power law in the number of particles, consistent with the Luttinger-liquid paradigm
and  the breakdown of Fermi-liquid theory. 
The variational Ansatz, however, predicts a finite value of $Z$, even in the thermodynamic limit. We also study how the presence of the impurity affects the density of the spin-up sea by calculating the pair correlation function. Subtracting the homogeneous background and integrating over all distances gives the charge $Q$. This charge turns out to grow continuously from 0 at zero coupling to 1 in the strong-coupling limit. The varational Ansatz predicts $Q=0$ at all couplings. So, although the variational Ansatz has been shown to be remarkably accurate for the energy and the effective mass, it fails even qualitatively when predicting $Z$ and the pair correlation function in the thermodynamic limit. 
\end{abstract}

\maketitle

\section{Introduction}
A  single mobile impurity coupled to a bath of particles is a fundamental many-body problem that appears in many different fields of physics. 
Typical examples are proton impurities in neutron stars~\cite{Nemeth_1968} or excitons in doped semiconductors~\cite{Sidler_2017}. 
Due to the interaction with the bath, the impurity is dressed with excitations of the bath and becomes a quasi-particle called polaron. 
Such quasi-particles were first described by Landau and Pekar in the context of an electron moving in a crystal lattice~\cite{Landau_1933,Pekar_1946,Landau_1948},
where the coupling to phonons renormalizes the characteristic properties of the bare electron. 
In the last decades, the creation and unprecedented control of ultracold atomic mixtures has initiated a very active research activity on polaron physics~\cite{Massignan_2014,Schmidt_2018,Scazza_2022,Parish_2023}. 
In these experiments polarons are created in atomic mixtures through population imbalance: the minority atoms form mobile impurities in a bath formed by the majority atoms. 
A key feature is the tunability of the two-body interaction between the impurity and the bath particles via Feshbach resonances. 
Baths consisting of bosonic~\cite{Hu_2016,Jorgensen_2016,Yan_2020,Skou_2021} and fermionic~\cite{Schirotzek_2009,Nascimbene_2009,Liao_2010,Kohstall_2012,Koschorreck_2012,Zhang_2012,Wenz_2013,Ong_2015,Cetina_2015,Cetina_2016,Scazza_2017,Mukherjee_2017,Yan_2019,Darkwah_2019,Ness_2020,Fritsche_2021} atoms, corresponding respectively to the so-called Bose and Fermi polarons, have both been realized.

In the Fermi polaron problem one considers a single spin-down impurity in the continuum interacting with an ideal gas of spin-up fermions through an attractive potential in the zero-range limit. In two (2D) and three (3D) spatial dimensions, a stable polaron quasi-particle is formed at weak coupling. This polaron consists of the impurity dressed with particle-hole excitations of the Fermi sea. At strong coupling, however, the ground state corresponds to a dimeron state. This is a bound pair of fermions of opposite spin dressed with particle-hole excitations of the Fermi sea. The value of the critical interaction strength marking the polaron-to-dimeron first-order transition was determined by diagrammatic Monte Carlo in 3D~\cite{polaron1,polaron2,Vlietinck_3D} and 2D~\cite{Vlietinck_2D}. These Monte Carlo simulations take a high number of particle-hole excitations into account. 
Remarkably, the basic ground-state properties of the system can be accurately approximated by a simple variational Ansatz where the Hilbert space for the excited states of the Fermi sea is restricted to have at most one or two particle-hole pairs~\cite{Chevy_2006, Combescot_2008}. Such Ans\"atze have been extensively used to study the properties of the Fermi polaron in 3D~\cite{Punk_2009,Mora_2009,Bruun_2010,Schmidt_2011,Mathy_2011,Trefzger_2012}
 and 2D~\cite{Parish_2011,Zollner_2011,Parish_2013,Cui_2020,Peng_2021}. 
 In one spatial dimension (1D), the model is exactly solvable by Bethe Ansatz. The exact solution shows no sharp transition but rather a smooth crossover from weak to strong coupling~\cite{McGuire_1966}.
The variational Ansatz has also been tested in 1D. Similar to higher spatial dimensions, the variational Ansatz 
 gives remarkably accurate values for the energy as well as the effective mass of the impurity in 1D~\cite{Giraud_2009_1D}. 
 For a recent review of theoretical and experimental studies of the properties of polarons formed by mobile impurities strongly interacting with quantum many-body systems, see~\cite{Massignan_2025}.

In this article we focus on two physical properties of the 1D Fermi polaron in the thermodynamic limit which we will calculate exactly and via a variational Ansatz. 
The first quantity is the quasi-particle residue, which is the overlap of the interacting state with the non-interacting one:
\begin{equation}
Z(N_{\uparrow}) = | \langle \psi     |  \psi_\textrm{NI} \rangle   |^2 \; ,
\label{eq:Z_def}
\end{equation}
with $|\psi \rangle$ and $|  \psi_\textrm{NI} \rangle$ respectively the normalized ground state of the interacting and non-interacting system  with zero total momentum. 
More specifically,  $ |  \psi_\textrm{NI} \rangle = \hat{c}^{\dagger}_{k = 0,\downarrow}    | \text{FS}_{N_{\uparrow}} \rangle $, with $ | \text{FS}_{N_{\uparrow}} \rangle $  the ideal Fermi sea of $N_{\uparrow}$ spin-up fermions and the operator $ \hat{c}^{\dagger}_{k,\sigma}$ creating
a spin-$\sigma$ fermion carrying momentum $k$. The spin-up particle number $N_{\uparrow}$ is taken odd so that the Fermi sea has zero total momentum. 
Alternatively, $Z(N_{\uparrow})$ can also be viewed as the  quasi-particle spectral weight, i.e.  the strength of the quasi-particle peak in the spectral function. It should remain finite in the thermodynamic limit if Fermi liquid theory is valid. Previous studies have mainly focused on the repulsive 1D polaron problem~\cite{Castella_1993, Levinsen_2015}. 
In a recent work~\cite{Liu_2025} Bethe Ansatz was used to obtain exact spectral properties of the one-dimensional Hubbard model with $N_\uparrow$ spin-up fermions and one spin-down impurity with attractive and repulsive on-site interaction.  In this article we study the attractive model in the continuum.

The second quantity is the charge $Q$ of the polaron, which is a measure for the number of excess fermions in the excitation cloud surrounding the impurity. To properly define $Q$, we first consider the density-density or pair correlation function 
\begin{equation}
g_2(x) =  \langle  \hat{n}_{\uparrow}(x)  \hat{n}_{\downarrow}(0)  \rangle \; ,
\end{equation}
with $ \hat{n}_{\sigma}(x)$ the number-density operator at position $x$ for spin-$\sigma$ fermions. 
The correlation function $g_2(x)$ is proportional to the probability of finding a spin-up fermion at position $x$ relative to the position of the spin-down impurity. 
Since we consider a single spin-down fermion, the homogeneous spin-down density is $n_{\downarrow} = 1/L$, with $L$ the length of the system. We define $\tilde{g}_2(x) = g_2(x)/n_{\downarrow}$.
From the correlation function one can define the charge of the quasi-particle~\cite{Chen_2018}: 
\begin{equation}
Q = \int dx \lim_{L\to \infty} \left(   \tilde{g}_2(x) - n_{\uparrow}     \right) \; ,
\label{eq:charge_def}
\end{equation}
where the thermodynamic limit ($L \to \infty$ and $N_{\uparrow} \to \infty$ while keeping $n_{\uparrow}=N_{\uparrow}/L$ fixed) should be taken before integration (otherwise the result would always be trivially zero). 
The presence of the impurity perturbes the homogeneous density $n_{\uparrow}$ of the spin-up Fermi sea due to the attractive contact interaction. 
This leads to a peak in $\tilde{g}_2(x)$ at $x=0$ followed by Friedel-like oscillations at finite distance. 
The charge $Q$ is obtained by subtracting the homogeneous background and integrating over distance (after the thermodynamic limit has been taken).

In Refs.~\cite{Massignan_2005,Massignan_2025} the total number $\Delta N$ of extra majority particles attracted to the impurity was calculated from thermodynamic arguments by requiring that the density of the majority particles far away from the impurity remains constant, leading to 
\begin{equation}\label{defDeltaN}
\Delta N  = - \frac{\partial E_P}{\partial E_F}\;.
\end{equation}
In Eq.(\ref{defDeltaN}) $E_P$ represents the polaron energy, which is defined as the change in the ground state energy of the system induced by the interaction, $E_P= E_\textrm{gs}(g)-E_\textrm{gs}(0)$.  
We will show numerically that for the 1D Fermi polaron $\Delta N$ coincides with the charge $Q$ defined in Eq.~(\ref{eq:charge_def}).

The correlation function could be obtained experimentally by measuring the positions of all particles and then averaging over many realisations. This has become possible within current quantum gas microscope experiments that have single-atom resolution~\cite{Feng_2025,Daix_2025}. 
In these experiments, one can directly measure the density-density correlation function, from which one could in principle extract the charge.

We will calculate $Z(N_{\uparrow})$ and $Q$ for the 1D Fermi polaron problem via Bethe Ansatz and variational Ansatz. 
Moreover, we will also provide additional diagrammatic Monte Carlo data for $Z(N_{\uparrow})$, validating the results obtained via Bethe Ansatz. 
These calculations show that $Z$ vanishes in the thermodynamic limit, consistent with Luttinger-liquid theory.
Calculation of the charge $Q$ for the mobile impurity in 1D will show that $Q$ is not a good quantum number in this case: $Q$ goes continuously from $0$ to $1$ upon increasing the attraction strength, quantifying the crossover from weak to strong coupling.
Moreover, the variational Ansatz will turn out be very inaccurate for both of these \emph{thermodynamic-limit} quantities.  
This is in stark contrast with the 3D Fermi polaron, where the variational Ansatz gives very accurate values of $Z$ from weak to strong coupling even in the thermodynamic limit~\cite{Vlietinck_3D}.

The article is organized as follows: in Section~\ref{sec:model} we briefly introduce the model, while different methods (Bethe Ansatz, diagrammatic Monte Carlo and variational Ansatz) for calculating the quasi-particle properties are presented in Section~\ref{sec:methods}. Results are shown and discussed in Section~\ref{sec:results}.

\section{Model}\label{sec:model}

We consider a system of $N_{\uparrow}$ spin-up fermions in one dimension interacting with a single spin-down fermion through an attractive contact interaction with strength $g<0$. 
All fermions have equal mass $m$. This is a special case of the Yang-Gaudin model~\cite{Yang_1967,Gaudin_1967} (for a recent review see ~\cite{Guan_RMP2013}). 
The spin-up fermion density is
$n_{\uparrow} = N_{\uparrow}/L = k_F/\pi$, with $k_F$ the Fermi wave vector of the spin-up Fermi sea.  
The model Hamiltonian is
\begin{equation}
\hat{H} =  \sum_{i=1}^{N_{\uparrow}+1}  \frac{\hat{p}_{i}^2}{2m}
+ g \sum_{i=1}^{N_{\uparrow}} \delta(\hat{x}_{i}-\hat{x}_{N_{\uparrow}+1}) \; ,
\label{eq:H_Gaudin}
\end{equation}
where we use the convention that the $(N_{\uparrow}+1)$-th coordinate refers to the spin-down impurity, so that  
 $\hat{x}_{i}$ and $\hat{p}_{i}$ are the position and momentum operators of the spin-up fermions for $1\leq i \leq N_{\uparrow}$,  and of the spin-down impurity if $i=N_{\uparrow}+1$.
We set $\hbar=1$. We also define a dimensionless interaction strength $\gamma = m g /n_{\uparrow} = -2\pi/(k_F a_{\text{1D}})$, with $a_{\text{1D}}$ the 1D scattering length.

\section{Methods}\label{sec:methods}

\subsection{Bethe Ansatz}

The 1D Yang-Gaudin model is integrable by Bethe Ansatz and
the corresponding exact solution for the polaron problem (\ref{eq:H_Gaudin}) was first obtained by McGuire \cite{McGuire_1966}. 
Here, we focus on calculating the quasi-particle residue $Z$  of the polaron at zero momentum, $k=0$  (see Eq.~(\ref{eq:Z_def})), 
 as well as the charge $Q$ defined in Eq.(\ref{eq:charge_def}). 
In this section, the ground state wavefunction $\psi$ and  its non-interacting counterpart $\psi_\textrm{NI}$ are not assumed to be normalized.
We then write $Z$ as the normalized overlap
\begin{equation}\label{Z}
	Z=\frac{|\langle\psi_\textrm{NI}|\psi\rangle|^2}{\langle\psi_\textrm{NI}|\psi_\textrm{NI}\rangle \langle\psi|\psi\rangle }\; .
\end{equation}	
The inner product between two generic wave functions $\psi^\prime$ and $\psi$  is  defined as 
\begin{eqnarray}\label{scalprod}
	\langle \psi^\prime|\psi\rangle
	& = & \int_0^L \left( \prod_{j=1}^{ N_{\uparrow}+1}dx_j \right) ~ \psi^\prime(x_1, \dots, x_{N_{\uparrow}}, x_{ N_{\uparrow}+1})^* \nonumber \\
	& & \times ~\psi(x_1, \dots, x_{N_{\uparrow}}, x_{ N_{\uparrow}+1})\; ,
\end{eqnarray}
with $x_{ N_{\uparrow}+1}$ the position of the spin-down impurity.

The many-body wave function is completely determined by a set of parameters, corresponding to the allowed fermion quasi-momenta $\{k_j\}$, with $j=1, \dots , N_{\uparrow}+1$, and the spin rapidity $\lambda$.  These quantities are
obtained by solving the following Bethe Ansatz equations \cite{Gaudin_1967,Yang_1967}
\begin{eqnarray}
	\frac{k_j-\lambda+i g^\prime}{k_j-\lambda-i g^\prime}&=&e^{i k_j L} \; , \label{BA1}
\end{eqnarray}
for $j=1,2,\ldots,N_{\uparrow}+1$, and 
\begin{eqnarray}
	\prod_{j=1}^{N_{\uparrow}+1} \frac{k_j-\lambda+i g^\prime}{k_j-\lambda-i g^\prime}&=&1\; ,\label{BA2}
\end{eqnarray}
where $g^\prime=mg/2$ and periodic boundary conditions have been assumed.

The ground state corresponds to the choice of the quasi-momenta  minimizing the total energy 
\begin{equation}\label{energy}
	E=\sum_{j=1}^{N_{\uparrow}+1} \frac{ k_j^2}{2m}\; .
\end{equation}
For $N_\uparrow$ odd, the spin rapidity $\lambda$ vanishes and the fermion quasi-momenta satisfy the relation $k_{2i}=-k_{2i-1}$. In particular,
two of the quasi-momenta  are purely imaginary, $k_{1,2} = \pm i \alpha$, 
with $\alpha=g^\prime$ in the  limit $L\rightarrow \infty$. These quasi-momenta give a negative
contribution to the ground state energy (\ref{energy}), corresponding to the two-body bound state energy in vacuum, $-E_B = -mg^2/4$. All the remaining quasi-momenta  $k_i$, with $i>2$, are instead real.

In the following we use the Takahashi representation \cite{Takahashi} of the many-body wave function, which allows for an efficient numerical computation of the spectral weight (\ref{Z}) even for $N_{\uparrow}$ of the order of few hundreds. The explicit expression is \cite{guanPRA2016,guanPRAerratum}: 
\begin{equation}\label{psi}
	\psi(x_1,\dots,x_{N_{\uparrow}},x_{N_{\uparrow}+1})=\sum_{\ell=1}^{N_{\uparrow}+1} \xi_\ell ~\textrm{det}(A^\ell) ~e^{i k_\ell x_{N_{\uparrow}+1}},
\end{equation}
where 
\begin{equation}
	\xi_\ell=\begin{cases} 
		1 & \textrm{if}\; N_{\uparrow}\; \textrm{even}  \;,\\
		(-1)^{\ell+1}  &\textrm{if}\; N_{\uparrow}\; \textrm{odd}\; . \\
	\end{cases}
\end{equation}
In Eq.~(\ref{psi}) $A^\ell$ are $N_{\uparrow}\times N_{\uparrow}$ matrices, whose 
entries are given by 
\begin{equation}\label{entries}
	(A^\ell)_{is}=[k_{r} -\lambda-i g^\prime \textrm{sign}(x_i-x_{N_{\uparrow}+1})]e^{i k_{r} x_i},
\end{equation}
with $i, s=1, \dots , N_{\uparrow}$ and 
$r=\ell +s$ if $\ell +s \leq N_\uparrow+1$ and $r=\ell +s-N_\uparrow-1$ otherwise. 
Next, we trade the coordinates of the spin-up particles for the corresponding distances  $y_i=x_i-x_{N_{\uparrow}+1}$ from the impurity, with $i=1,\dots,N_{\uparrow}$, and  restrict to $0\leq y_i \leq L$ without loss of generality. The wave function (\ref{psi}) then takes the factorized form
\begin{equation}\label{F}
		\psi(x_1,\dots,x_{N_{\uparrow}+1})=F(y_1,\dots,y_{N_{\uparrow}}) e^{i \sum_{j=1}^{N_{\uparrow}+1} k_j x_{N_{\uparrow}+1}},
\end{equation}
where the function $F$ is defined as
\begin{equation}\label{psi2}
	F(y_1,\dots,y_{N_{\uparrow}})=\sum_{\ell=1}^{N_{\uparrow}+1} \xi_\ell ~\textrm{det}(\tilde A^\ell)\; ,
\end{equation}
with
\begin{equation}\label{entries2}
	(\tilde A^\ell)_{is}=(k_{r} -\lambda-i g^\prime)e^{i k_{r} y_i} \; .
\end{equation}

Notice that  $F$ represents the many-body wave function in the reference frame of the impurity. 
The rhs of Eq.~(\ref{psi2}) corresponds to the expansion
by minors of a determinant of a $(N_\uparrow+1)\times 
(N_\uparrow +1)$ matrix, whose first row has all entries equal to $1$:
\begin{equation}\label{detNplus1}
	\begin{vmatrix}
	     1 & \dots & 1\\
		(k_1 -\lambda-i g^\prime)e^{i k_1 y_1}   & \dots & (k_{N_\uparrow+1} -\lambda-i g^\prime)e^{i k_{N_\uparrow+1} y_1} \\
		\vdots & \ddots & \vdots\\
		(k_1 -\lambda-i g^\prime)e^{i k_1 y_{N_\uparrow}} 	 &\dots & (k_{N_\uparrow+1} -\lambda-i g^\prime)e^{i k_{N_\uparrow+1} y_{N_\uparrow}} 	
	\end{vmatrix}.
\end{equation}

The  determinant in Eq. (\ref{detNplus1}) can be  simplified by replacing the first $N_\uparrow$ columns by their differences with respect to the last column, implying that all entries in the first row become $0$, except for the last element. 
As a result, the function $F$ in Eq.~(\ref{psi2}) can be expressed as a single Slater determinant 
\begin{equation}\label{FF}
F(y_1,\dots,y_{N_\uparrow})=  \begin{vmatrix}
		\varphi_1(y_1)    & \dots & \varphi_{N_\uparrow}(y_1) \\
		\vdots & \ddots & \vdots\\
		\varphi_1(y_{N_\uparrow})	 &\dots & \varphi_{N_\uparrow}(y_{N_\uparrow})	
	\end{vmatrix},  
\end{equation}
with the single-particle functions $\varphi_i$
being  defined as 
\begin{equation}\label{sp}
\varphi_i(y)=(k_i -\lambda-i g^\prime)e^{i k_i y}- (k_{N_\uparrow+1} -\lambda-i g^\prime)e^{i k_{N_\uparrow+1} y}, 	
\end{equation}
 with $i=1,\dots,N_{\uparrow}$. 
 
To compute the square norm $\langle\psi|\psi\rangle$ of the many-body wave function, we insert Eqs. (\ref{F}) and (\ref{FF}) in Eq. (\ref{scalprod}), with the coordinate variables $x_i$ being replaced by the distance variables $y_i$.
The integration over the impurity coordinate is straightforward, $\int_0^L dx_{N_{\uparrow}+1}=L$. 
For the integration over the remaining variables, we make use of the fact  that the inner product between two  Slater determinants can be recast as a Slater determinant of 
1D integrals,   thanks to the general identity
\begin{eqnarray}
	&\int_0^L& \prod_{j=1}^{M}dy_j 
	\begin{vmatrix}
		\phi_1^*(y_1)    & \dots & \phi_{M}^*(y_1) \\
		\vdots & \ddots & \vdots\\
		\phi_1^*(y_{M})	 &\dots & \phi_{M}^*(y_{M})	
	\end{vmatrix}  
	\begin{vmatrix}
		\varphi_1(y_1)    & \dots & \varphi_{M}(y_1) \\
		\vdots & \ddots & \vdots\\
		\varphi_1(y_{M})	 &\dots & \varphi_{M}(y_{M})	
	\end{vmatrix}	
		\nonumber \\
   &=&  M!
	\begin{vmatrix}
		\int_0^L dy	\phi_1^*(y)\varphi_1(y)    & \dots & 	\int_0^L dy	\phi_1^*(y)\varphi_M(y)  \\
		\vdots & \ddots & \vdots\\
		\int_0^L dy  \phi_M^*(y) \varphi_1(y)	 &\dots & \int_0^L dy  \phi_M^*(y) \varphi_M(y)	\\
	\end{vmatrix}.
	\label{identity}
\end{eqnarray}
We first use Eq.~(\ref{identity}) with $M=N_\uparrow$ and $\phi_j=\varphi_j$. 
The 1D overlaps can be computed analytically from Eq.~(\ref{sp}) yielding
\begin{eqnarray}\label{eqlong}
&&\int_0^L  dy	\varphi_i^*(y) \varphi_j(y) \\
=&&(k_i^* -\lambda+i g^\prime)
(k_j -\lambda-i g^\prime)  R_L(k_j-k_i^*) \nonumber \\
-&& (k_i^* -\lambda+i g^\prime)
(k_{N_\uparrow+1} -\lambda-i g^\prime) R_L(k_{N_\uparrow+1}-k_i^*) \nonumber \\
-&& (k_{N_\uparrow+1}^* -\lambda+i g^\prime)
(k_j -\lambda-i g^\prime) R_L(k_j-k_{N_\uparrow+1}^*) \nonumber \\
+&& (k_{N_\uparrow+1}^* -\lambda+i g^\prime)
(k_{N_\uparrow+1} -\lambda-i g^\prime) R_L(k_{N_\uparrow+1}-k_{N_\uparrow+1}^*) \nonumber, %\\
\end{eqnarray}
where the function $R_L(p)$ is defined as
\begin{equation}\label{defR}
R_L(p)=\int_0^L e^{i p y}dy=
\begin{cases} 
      \frac{(e^{i pL}-1)}{ip} & \textrm{if}\; p\neq 0 \\
      L & \textrm{if}\; p=0 \; .\\
   \end{cases}
\end{equation}
The rhs of Eq.~(\ref{eqlong}) can be further simplified whenever the argument $p$ of the function $R_L$ is nonzero, by making use of Eq.~(\ref{BA1}). 
For instance, by taking into account that
\begin{eqnarray}
&&(k_i^* -\lambda+i g^\prime) e^{-i k_i^*L} (k_j -\lambda-i g^\prime)e^{i k_jL}  \\
&&=(k_i-\lambda+i g^\prime)^* (k_j -\lambda+i g^\prime)\nonumber \\
&&=(k_i^*-\lambda-i g^\prime)(k_j -\lambda+i g^\prime) \; ,
\end{eqnarray}
 the first term in the rhs of Eq.~(\ref{eqlong}) for $k_j-k_i^*\neq 0$ can be recast as 
\begin{equation}
\frac{(D-i g^\prime)(E+ig^\prime)-(D+ig^\prime)(E-ig^\prime)}{i(E-D)}=-2g^\prime \; ,   
\end{equation}
with $D=k_i^* -\lambda$ and $E=k_j -\lambda$. In contrast, for
$k_j-k_i^*= 0$ the same term reduces to $(k_j-\lambda)^2+g^\prime{}^2$.
Similar considerations apply also to the other terms in the rhs of Eq.~(\ref{eqlong}). 
In particular, for $N_\uparrow >1$ the second and the third terms yield a contribution of $+2g^\prime$ each, while the last term reduces to  $(k_{N_\uparrow+1}-\lambda)^2+g^\prime{}^2$, because $k_{N_\uparrow+1}$ is real.

To reduce overflow or underflow issues in numerics,  we find it convenient to rescale the matrix of 1D overlaps in Eq.~(\ref{identity})
by dividing all its elements by $L$, so that the square norm of the wavefunction takes the final form
\begin{equation}\label{normpsi}
\langle\psi|\psi\rangle=N_\uparrow !\; L^{N_\uparrow +1} \textrm{det}(S)\;,
\end{equation}
where the matrix $S$ has the form 
\begin{equation}\label{S}
S=u_{N_\uparrow +1} J+
    \begin{pmatrix}
     & u_1 \\
    u_2 &  \\
    & &  u_3  \\
    & &  & u_4  \\
    & &  &  &\ddots &\\
    & &   & & & u_N
 \end{pmatrix}\;,   
\end{equation}
with $J$ being a constant matrix with all elements equal to $1$ and
\begin{equation}
u_i=(k_i-\lambda)^2+g^{\prime 2} +2g^\prime/L\;.   
\end{equation} 
Notice that in Eq.(\ref{S}) we use the convention that omitted matrix entries are zero.

The  wave function of noninteracting spin-1/2 fermions
is the product of the Slater determinants of the two spin components. 
For our $N_\uparrow +1$ particles system, this reduces to 
\begin{equation}\label{psiNI}
\psi_{NI}(x_1,..,x_{N_\uparrow +1})=	\begin{vmatrix}
	   e^{i q_2 x_1} & \dots & e^{i q_{N\uparrow+1}x_1} \\
		\vdots & \ddots & \vdots\\
		e^{i q_2 x_{N_\uparrow}}  & \dots & e^{i q_{N\uparrow+1}x_{N_\uparrow}} \\	
	\end{vmatrix}
	e^{i q_1 x_{N_\uparrow +1}} \; , 
\end{equation}
where the fermion quasi-momenta $q_j$ satisfy the condition 
$e^{i q_j L}=1$, as follows from Eq.~(\ref{BA1}) by setting  $g^\prime=0$. The ground state wavefunction corresponds to the choice $q_1=0$ and $q_{i}=2\pi n/L$ for $i>1$, with $n=0,\pm 1,\pm 2, \dots$.

Written in terms of the distance variables, Eq.~(\ref{psiNI}) takes the form 
\begin{equation}
\psi_{NI}(x_1,..,x_{N_\uparrow +1})=F_{NI}(y_1,\dots,y_{N_{\uparrow}}) e^{i \sum_{j=1}^{N_\uparrow+1} q_j x_{N_\uparrow +1}}\; ,
\end{equation}
with
\begin{equation}\label{psiNI2}	
F_{NI}(y_1,\dots,y_{N_{\uparrow}})=\begin{vmatrix}
	   e^{i q_2 y_1} & \dots & e^{i q_{N\uparrow+1}y_1} \\
		\vdots & \ddots & \vdots\\
		e^{i q_2 y_{N_\uparrow}}  & \dots & e^{i q_{N\uparrow+1}y_{N_\uparrow}} \\	
	\end{vmatrix} \; .
\end{equation}
The square norm of the noninteracting wave function can 
be easily calculated from Eq.~(\ref{identity}). The functions $e^{i q_j y}$, with $j=2,\dots,N_\uparrow+1$, are periodic in the interval $[0,L]$ and orthogonal to each other, 
$\int_0^Le^{i (q_j-q_s)y}dy=L\delta_{js}$. 
Taking into account the integration over the impurity coordinate, we  obtain
\begin{equation}\label{normpsiNI}
\langle\psi_{NI}|\psi_{NI}\rangle=N_\uparrow !\; L^{N_\uparrow +1}\; .    
\end{equation}

We now turn to the computation of the scalar product  between the two many-body wave functions.
 The integration over the the impurity variable is nonzero provided the total momentum is equal for the two wave functions, that is $\sum_{j=1}^{N_\uparrow+1} k_j=\sum_{j=1}^{N_\uparrow+1} q_j$. For the integration over the remaining variables, we substitute Eq.~(\ref{sp}) in Eq.~(\ref{identity}), with $\phi_i(y)=e^{i q_i y}$, and evaluate the 1D overlaps:
\begin{eqnarray}\label{overlaps}
\!\!\int_0^L e^{-i q_i y}\varphi_j(y) dy
=(k_j-\lambda-i g^\prime)R_L(k_j-q_i)\nonumber \\ 
-(k_{N_\uparrow+1} -\lambda-i g^\prime) R_L(k_{N_\uparrow+1} -q_i)\; .
\end{eqnarray}
Making again use of Eq.~(\ref{BA1}), Eq.~(\ref{overlaps}) reduces to 
\begin{equation}
   \int_0^L e^{-i q_i y}\varphi_j(y) dy=\frac{2g^\prime}{k_j-q_i}- \frac{2g^\prime}{k_{N_\uparrow+1}-q_i}\; ,
\end{equation}
implying that 
\begin{equation}\label{over}
\langle\psi_{NI}|\psi\rangle=N_\uparrow !\; L^{N_\uparrow +1} \textrm{det}(V)\; ,    
\end{equation}
where $V$ is a $N_\uparrow \times N_\uparrow$ matrix, whose elements are defined as
\begin{equation}\label{over2}
   V_{ij}=\frac{2g^\prime}{L} \left ( \frac{1}{k_j-q_i}- \frac{1}{k_{N_\uparrow+1}-q_i}\right)\; .
\end{equation}
By substituting Eqs. (\ref{normpsi}), (\ref{normpsiNI}) and (\ref{over}) into
Eq.~(\ref{Z}), we find that the spectral weight $Z$ takes the final form  
\begin{equation}
Z=\frac{|\textrm{det}(V)|^2}{\textrm{det}(S)}\; ,    
\end{equation}
which sets the basis for our numerics.

To calculate the charge $Q$ from Eq.~(\ref{eq:charge_def}) we start from the pair correlation function 
\begin{align}
& g_2(x-x^\prime) 
 =    \langle  \hat{n}_{\uparrow}(x)  \hat{n}_{\downarrow}(x^\prime)  \rangle  \nonumber \\
& =  \frac{N_{\uparrow}}{\langle\psi|\psi\rangle}
\int_0^L \left( \prod_{j=2}^{ N_{\uparrow}}dx_j \right) ~ \psi(x, x_2,\dots, x_{N_{\uparrow}}, x^\prime)^* \nonumber \\
&	 \times ~\psi(x, x_2,\dots, x_{N_{\uparrow}}, x^\prime)\; ,
\end{align}
with $\psi$ the Bethe Ansatz solution. 
Following Refs.~\cite{McGuire_1966,Giraud_thesis} the thermodynamic-limit $\tilde{g}_2 \equiv g_2/n_{\downarrow}$ is given by
\begin{equation}
\tilde{g}_2(x) =  n_\uparrow - \frac{n_\uparrow\gamma |h(x)|^2}{1-\frac{2}{\pi}\arctan(\gamma/2\pi)} \; ,
\label{eq:g2_Bethe}
\end{equation}
with\begin{equation}
h(x) = e^{k_F |x| \gamma/2\pi } \theta(-\gamma) - \frac{i}{2\pi} \int_{-1}^1 \frac{e^{-iuk_F|x|}}{u-i\gamma/2\pi}du \; .
\label{eq:h}
\end{equation}
The integral in Eq.~(\ref{eq:h}) can easily be evaluated numerically. We then numerically integrate 
$\tilde{g}_2(x)-n_\uparrow$ over position $x$ to give the charge $Q$.
Note that the first term in Eq.~(\ref{eq:h}) is due to the two-body bound state in vacuum, which is always present in 1D for an attractive potential, $\gamma<0$.
In the strong-coupling limit, we can drop the second term in   Eq.~(\ref{eq:h}), giving the leading contribution $\tilde{g}_2(x)-n_\uparrow \simeq - n_{\uparrow} \gamma \exp(n_{\uparrow}|x|\gamma)/2$. Integration over $x$ gives charge $Q=1$ in this limit.

The total number $\Delta N$ of extra majority fermions surrounding the impurity defined in Eq.~(\ref{defDeltaN})
can also be calculated by Bethe ansatz. For a finite system
 the polaron energy is readily computed as
\begin{equation}\label{EpN}
    E_P(N_\uparrow)=\sum_{i=1}^{N_\uparrow+1} \left (\frac{k_i^2}{2m} -\frac{q_i^2}{2m} \right)=-E_B+\sum_{i=3}^{N_\uparrow+1} \left (\frac{k_i^2}{2m} -\frac{q_i^2}{2m} \right).
\end{equation}
We can trade the real fermion quasimomenta $k_{i}$, with $i>2$, with the corresponding phase shifts defined through the relation
\begin{equation}
 k_i=q_i-\frac{2}{L}\delta_i.  
\end{equation}  
For infinite systems, the polaron energy can be written as
an integral over the phase shifts of the occupied states:
\begin{eqnarray}
E_P&\simeq -&E_B-\sum_{i=3}^{N_\uparrow+1} \frac{2}{mL} q_i \delta_i \nonumber\\
&\simeq &-E_B -\frac{2}{\pi}\int_0^{k_F}\frac{q}{m}\delta_s(q) dq \; , \label{Fumi}
\end{eqnarray}
in agreement with Fumi's theorem. Here we have labelled the continuous phase shifts as $\delta_s(q)$, with the
underscript $s$ referring to the s-wave scattering. 
Taking into account that $E_F=k_F^2/(2m)$, from Eq.~(\ref{Fumi}) we
find
\begin{equation}\label{DeltaN}
\Delta N =\frac{2}{\pi} \delta_s(k_F)\;,
\end{equation}
showing that $\Delta N$  depends solely on the phase shift at the Fermi edge. The polaron energy for infinite systems is given by \cite{McGuire_1966}
\begin{equation}\label{Ep}
E_P=-E_B+\frac{2}{\pi} E_F \left( y-\frac{\pi}{2}y^2+(1+y^2)\arctan y  \right )\;,  
\end{equation}
with $y=mg/(2k_F)=\gamma/(2\pi)$. The exact solution 
for the repulsive case \cite{McguireJMP1965}  shows that
the polaron energy is still given by Eq.~(\ref{Ep}) with 
  $E_B=0$ (note that there is no bound state for $g>0$). Combining
 Eq.~(\ref{Fumi}) with Eq.~(\ref{Ep}) yields \cite{McGuire_1966} 
\begin{equation} \label{deltaN}
	\delta_s(k_F)=-\frac{\pi}{2} \frac{\partial E_P}{\partial E_F}=-\arctan\left(\frac{m g}{2k_F}\right)\;,
\end{equation}
holding irrespective of the sign of the interaction strength.

\subsection{Diagrammatic Monte Carlo}

In this section we briefly discuss the diagrammatic Monte Carlo (DiagMC) method. We will use this numerical method to determine the 
quasi-particle residue $Z$ of the $1\text{D}$ Fermi polaron and cross-validate the values obtained via Bethe Ansatz.

The central object calculated here by the DiagMC algorithm is  the single-particle propagator at zero momentum and imaginary time $\tau >0$:
\begin{eqnarray}
G_{\downarrow}(k=0,\tau) =  - \bra{\text{FS}_{N_{\uparrow}}} 
\hat{c}^{\phantom{\dagger}}_{0,\downarrow}(\tau) \hat{c}^{\dagger}_{0,\downarrow}(0)\ket{  \text{FS}_{N_{\uparrow}}},
\label{eq:Gdef}
\end{eqnarray} 
where the creation/annihilation operators are written in the imaginary-time Heisenberg picture, $\hat{c}^{\phantom{\dagger}}_{k,\downarrow}(\tau) = e^{\hat{K}\tau} \hat{c}_{k,\downarrow}  e^{-\hat{K}\tau}$ with $\hat{K}=\hat{H} -  \sum\limits_\sigma \mu_{\sigma} \hat{N}_{\sigma}$, and with $\mu_{\sigma}$ the chemical potentials for both spin components. The $\hat{N}_{\sigma}$ are the spin-$\sigma$ number operators. The ground-state polaron residue and energy can be extracted from the asymptotic large-$\tau$ behaviour of $G_{\downarrow}(0,\tau)$: 
\begin{equation}
G_{\downarrow}(0,\tau) \overset{\tau\to\infty}{=} - Z ~ e^{-(E_P-\mu_\downarrow)\tau} \; ,
\end{equation}
with the polaron energy $E_P = E - E_{\text{FS}}$ 
being the energy difference of the system with and without spin-down impurity. Here, $E_{FS}$ is the energy of the non-interacting Fermi sea: $\hat{H} | \text{FS}_{N_{\uparrow}} \rangle = E_{\text{FS}} | \text{FS}_{N_{\uparrow}} \rangle$.

We use a polaron determinant (PDet) algorithm for evaluating the diagrammatic expansion of $G_{\downarrow}(k,\tau)$ in powers of the bare coupling~\cite{pdet,Pascual_2024}. 
We performed simulations for the 1D attractive Hubbard model using the algorithm of Ref.~\cite{Pascual_2024}. The Hubbard Hamiltonian is given by
\begin{eqnarray}\label{eq: Hamiltonian}
\hat{H}  =   -t \sum_{\sigma=\uparrow,\downarrow}\sum_{i} \left( \hat{a}^{\dagger}_{i+1,\sigma}
\hat{a}^{\phantom{\dagger}}_{i,\sigma} + h.c. \right) + U \sum_{i} \hat{n}_{i,\uparrow} \hat{n}_{i,\downarrow}   \; , \; \; 
\end{eqnarray}
where $\hat{a}^{\dagger}_{i,\sigma}$ creates a spin-$\sigma$ fermion on site $i$, and $\hat{n}_{i,\sigma} = \hat{a}^{\dagger}_{i,\sigma}\hat{a}^{\phantom{\dagger}}_{i,\sigma}$ the on-site number operator. 
The first term describes hopping of fermions between nearest-neighbor sites with amplitude $t$,  while the second term describes on-site attraction with amplitude $U<0$.
We assume periodic boundary conditions and we set the lattice spacing to $b$. For the lattice model, one has $\gamma = U/(2\nu t)$ with $\nu=N_{\uparrow}/N=n_{\uparrow} b$ 
the spin-up filling fraction of the lattice with $N$ sites. In order to retrieve the properties of the Yang-Gaudin model, we take the $\nu \to 0$ limit at fixed $\gamma$ numerically. 
Note that, alternatively, a PDet algorithm could also be directly developed  for the Gaudin model (\ref{eq:H_Gaudin}) for an expansion in powers of $g$, 
since the Dirac-delta interaction potential causes no ultraviolet divergencies in 1D, in contrast to higher dimensions where a regularisation procedure is required.

The PDet algorithm works in position-imaginary time representation, where the sum of all diagram topologies for a given set of space-time coordinates of the interaction vertices is given by a single determinant~\cite{Rubtsov_2003,Rubtsov_2004,Burovski_2004}. For the polaron problem, 
all diagrams generated in this way are automatically connected~\cite{pdet}.  The expansion of the polaron propagator in powers of $U$ is written as  
\begin{eqnarray}
    G_{\downarrow}(X) & =  G_{\downarrow}^{0}(X) + \sum\limits_{n=1}^{\infty}
    \int_{X_1}\ldots \int_{X_n} 
   U^n G_{\downarrow}^{0}(X_1)   \nonumber \\
  &   G_{\downarrow}^{0}(X_2-X_1) \cdots G_{\downarrow}^{0}(X-X_n)  ~\det(\mathcal{M}^{(n)}) \; ,
\label{eq:Gexpansion}
\end{eqnarray}
with $\mathcal{M}^{(n)}$ an $n\times n$ matrix with elements $\mathcal{M}^{(n)}_{i,j} = G^{0}_{\uparrow}(X_i-X_j)$, with $i,j=1,\ldots,n$, where $n$ is the diagrammatic order. 
We have used the notation $X = (x,\tau)$ for position-imaginary-time coordinates and $\int_X := \sum\limits_{x}\int_0^{+\infty} d\tau$. 
The $G^{0}_{\sigma}$ are the non-interacting spin-$\sigma$ single-particle propagators.  The different terms in Eq.~(\ref{eq:Gexpansion}) are evaluated 
stochastically by the Metropolis algorithm. 

The data presented in Section~\ref{sec:results} is obtained for finite and fixed $N_{\uparrow}$.  The total number of lattice sites $N$ was increased while adjusting $U/t$ such that $\gamma$ is fixed. We have checked that our values of $N$ are large enough for all lattice effects to be within our statistical error bars.

\subsection{Variational Ansatz}

A powerful variational Ansatz for the Fermi polaron problem is obtained by restricting the Hilbert space for the excited states of the Fermi sea to have at most one or two particle-hole pairs~\cite{Chevy_2006, Combescot_2008}. Here, we consider the simplest Ansatz of maximally one particle-hole excitation. 
The variational Ansatz provides a very good approximation to the polaron energy and effective mass for the whole range of the interaction strength~\cite{Giraud_2009_1D}.
Moreover, in the strong-coupling limit the leading contribution of the Ansatz to the polaron energy correctly produces the two-body binding energy in vacuum, $E_B = mg^2/4$~\cite{Giraud_2009_1D}.   
The variational Ansatz for the polaron state with total momentum zero reads
\begin{equation}
|P\rangle  =  \Bigg[   
\alpha_0\hat{c}^{\dag}_{0,\downarrow}  +  \frac{1}{L}\sum_{k,q}{}^{'} 
\alpha_{k,q} \hat{c}^{\dag}_{q-k,\downarrow}\:
\hat{c}^{\dag}_{k,\uparrow}
\hat{c}_{q,\uparrow}
   \Bigg] |\text{FS}_{N_{\uparrow}}\rangle,  \;   \label{eq:Ansatz_Polaron}
\end{equation}
with normalisation condition $|\alpha_0|^2 + \sum{}^{'}_{k,q} |\alpha_{k,q}|^2/L^2 =1$. 
The prime in the sum
indicates that all $q$ are below the Fermi level of spin-$\uparrow$ fermions and all $k$ are above. The quasi-particle residue is given by $Z(N_{\uparrow}) = |\alpha_0|^2$.

In addition to the Ansatz~(\ref{eq:Ansatz_Polaron})  a variational Ansatz for the dimeron state has been proposed~\cite{Punk_2009,Combescot_2009,Mora_2009,Parish_2011,Parish_2013,Cui_2020,Peng_2021}.  Including again up to one particle-hole excitation, the dimeron Ansatz is
\begin{eqnarray} 
|M\rangle & = & \Bigg[
\frac{1}{\sqrt{L}}\sum_{k}{}^{'}\xi_{k}\hat{c}^{\dag}_{-k,\downarrow}\hat{c}^{\dag}_{k,\uparrow}  +  \frac{1}{L^{3/2}}\sum_{k,k',q}{}^{'}
\xi_{k,k',q}  \nonumber \\ & &
 \:\hat{c}^{\dag}_{q-k'-k,\downarrow}\:
\hat{c}^{\dag}_{k,\uparrow} \hat{c}^{\dag}_{k',\uparrow}
\hat{c}_{q,\uparrow}
\Bigg] |\text{FS}_{N_{\uparrow}-1}\rangle  \; , \label{eq:Ansatz_Dimeron}
\end{eqnarray}
with normalisation condition
\begin{equation}
\frac{1}{L}\sum_{k}    |\xi_{k}|^2 + \frac{2}{L^3} \sum_{k,k',q}{}^{'} |\xi_{k,k',q}|^2 = 1 \; .
\end{equation}
The prime in the sum again indicates that all $q$ are below the Fermi level and all $k, k'$ are above. In what follows we drop the prime to simplify the notations. 
For the 1D Fermi polaron, the polaron Ansatz (\ref{eq:Ansatz_Polaron}) always gives a lower energy than the dimeron Ansatz (\ref{eq:Ansatz_Dimeron}) at all coupling~\cite{Giraud_2009_1D,Giraud_thesis}. 

Evaluating $\tilde{g}_2(x)$ for the Ansatz $|P\rangle$ one finds
\begin{eqnarray}
\tilde{g}_2(x) & = & n_{\uparrow} + \frac{1}{L^3} \sum_{k,k',q} \alpha^*_{k,q} \alpha_{k',q} e^{-i(k-k')  x} \nonumber \\ & & 
- \frac{1}{L^3} \sum_{k,q,q'} \alpha^*_{k,q} \alpha_{k,q'} e^{-i(q'-q)  x} \nonumber \\
& & + \frac{1}{L^2}  \sum_{k,q}  \left( \alpha_0^*  \alpha_{k,q} e^{-i(q-k)  x} +h.c. \right) \; .
\label{eq:g2_P3}
\end{eqnarray}
Taking the thermodynamic limit, assuming that $\alpha_{k,q}$ is a smooth function and integrating over position gives charge $Q=0$. 
We have also checked this numerically by calculating $\tilde{g}_2(x)$ in the limit of large $L$ and $N_{\uparrow}$ at fixed $n_\uparrow = N_\uparrow/L$ and then integrating over $x$ (see also section~\ref{subsec:charge}). 
Similarly for the Ansatz $|M\rangle$ one finds
\begin{eqnarray}
\tilde{g}_2( x) & = & \frac{N_{\uparrow}-1}{ L}   + \frac{1}{ L^2} \sum_{k,k'}   \xi^*_{k}  \xi_{k'} e^{-i(k-k')   x} \nonumber \\
& &  + \frac{2}{ L^{3}} \sum_{k,k',q} \left(\xi_{k}^* \xi_{k,k',q} e^{-i(q-k')   x} +h.c.\right) \nonumber \\
& & - \frac{2}{ L^4} \sum_{k,k',q,q'} \xi_{k,k',q}^* \xi_{k,k',q'} e^{-i(q'-q)   x}  \nonumber \\
& & + \frac{4}{ L^4} \sum_{k,k',q,k''} \xi_{k,k',q}^* \xi_{k,k'',q} e^{-i(k'-k'')   x}
\; ,
\label{eq:g2_M4}
\end{eqnarray}
from which one gets, after taking the thermodynamic limit and integrating over position, charge $Q=1$. Here we assume again that $\xi_{k}$ and $\xi_{k,k',q}$ are smooth functions in the thermodynamic limit.

The fact that the polaron Ansatz gives $Q=0$ and the dimeron Ansatz gives $Q=1$ is straightforwardly generalized to 2D and 3D.  There, $Q$ will jump at the first-order phase transition between polaron and dimeron state.  The spatial structure of the pair correlation function was studied for the polaronic branch in 3D in Ref.~\cite{Trefzger_2012} within the one particle-hole variational Ansatz, showing that the polaron is a spatially extended object such that $Q=0$.

\section{Results and discussion} \label{sec:results}

\subsection{Quasi-particle residue and polaron energy}

\begin{figure}[!]
	\centering
	\includegraphics[width=.95\columnwidth]{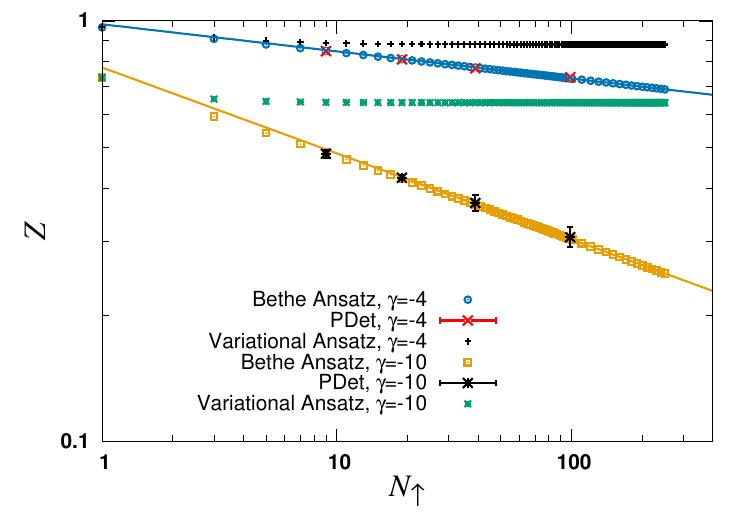}
	\caption{The quasi-particle residue $Z$ as function of the number $N_{\uparrow}$ of spin-up fermions for two interaction strengths, $\gamma=-4$ and $\gamma=-10$. The open symbols are obtained via Bethe Ansatz. The red and black symbols with error bars are results of diagrammatic Monte Carlo simulation with the PDet algorithm. Both are in excellent agreement. The full lines are power law fits to $Z = a N_{\uparrow}^{-\theta}$ [ with $a = 0.982$ and $\theta=0.0644$ for $\gamma=-4$,  and $a= 0.775$ and $\theta=0.2040$ for $\gamma=-10$ ]. Results obtained with the variational Ansatz are also shown. Here, the values of $Z$ quickly saturate to a constant upon increasing $N_{\uparrow}$.}
    \label{fig:Zfactor_N}
\end{figure}

Fig.~\ref{fig:Zfactor_N} shows the residue $Z$ as function of the number $N_{\uparrow}$ of fermions in the Fermi sea for interaction strengths $\gamma=-4$ and $\gamma=-10$.
The Bethe Ansatz data (open symbols) clearly shows a power law decay with $N_{\uparrow}$. 
The polaron's vanishing $Z$ in the thermodynamic limit is an example of Anderson's orthogonality catastrophe~\cite{Anderson_1967,Anderson_1967_2},
implying a breakdown of perturbation theory and non-Fermi liquid behavior. The physics of the 1D model is well described by Luttinger-liquid theory and the Fermi liquid theory is unstable due to divergencies in perturbation theory.

 The solid in lines in Fig.~\ref{fig:Zfactor_N}  are  fits of the data tails  with a power law $Z=a N_\uparrow^{-\theta}$, where $a$ and $\theta$ are fitting parameters. The obtained value for the exponent $\theta$ coincides (within error bars) with the analytical expression previously found  for repulsive interactions in Ref.~\cite{Castella_1993}, namely
 \begin{equation}\label{theta}
 \theta=\frac{2\delta_s(k_F)^2}{\pi^2}\; , 
 \end{equation}
 with the phase shift $\delta_s(k_F)$ given in Eq.~(\ref{deltaN}). 
  Our numerics suggests that Eq.~(\ref{theta}) is also verified for the attractive case, implying that the exponent $\theta$  is independent of the sign of the coupling constant $g$. Notice that Anderson's orthogonality catastrophe is driven by the single-particle states close to the Fermi edge, sharing approximately the same value of the phase shift (up to a sign). 
 
 Equation (\ref{theta}) was first derived by Anderson ~\cite{Anderson_1967_2} for the case of free-fermion scattering against a static  potential, corresponding to an impurity with infinite mass. Later,  
Affleck and Ludwig \cite{Affleck_JPA1994,AffleckProc1997} used boundary conformal field theory
arguments to confirm Eq. (\ref{theta}) and showed  that the 
exponent $\theta$ is also related to the 
finite-size correction $\Delta E=o(N_{\uparrow}^{-1})$ to the change in the ground state energy of the system induced by the scattering potential   through 
\begin{equation}\label{DeltaE}
\theta=\frac{\Delta E}{2 E_F}{N_\uparrow}\; .
\end{equation}

\begin{figure}[!]
	\centering
	\includegraphics[width=.95\columnwidth]{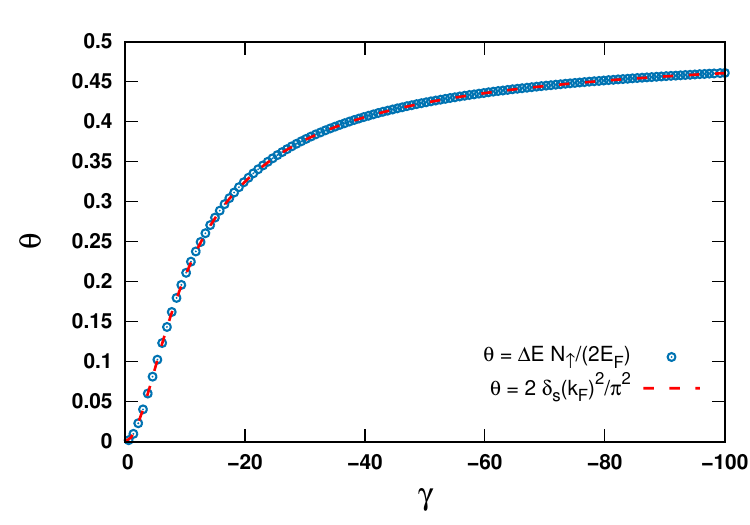}
	\caption{The exponent $\theta$, describing the Anderson orthogonality catastrophe, as a function of the interaction strength $\gamma$. The blue circles correspond
    to the numerical data obtained from  the finite-size correction of the polaron energy calculated with $N_\uparrow=99$, see Eq.~(\ref{DeltaE}).
	The dashed line corresponds to the analytical result,
   $\theta(\gamma)=2 \arctan^2(\gamma/(2\pi))/\pi^2$, obtained 
   from Eqs. (\ref{theta}) and (\ref{deltaN}). 
}
    \label{fig:affleck}
\end{figure}

 It is interesting to verify whether Eq.~(\ref{DeltaE}) agrees with Eq.~(\ref{theta}) also for our specific system, where the scattering process originates from interactions with a mobile impurity, having  the same mass of the fermions in the bath. 
To this end, we compute the finite-size correction to the polaron energy as $\Delta E=E_P(N_\uparrow)-E_P$, where 
$E_P(N_\uparrow)$ and its thermodynamic limit counterpart $E_P$
are given in Eqs. (\ref{EpN}) and (\ref{Ep}), respectively.
In Fig.\ref{fig:affleck} we plot the value of $\theta$, extracted from Eq.~(\ref{DeltaE}), as a function of the interaction parameter $\gamma$ (blue circles). The obtained results are in excellent numerical agreement with Eq. (\ref{theta})  (red-dashed line). 
This fact can be understood from Eq.~(\ref{FF}), showing that the many-body function $F$ in the impurity frame can be written as a Slater determinant of effective single-particle functions of the distance variables, 
in analogy to the many-body wave function  of noninteracting spinless fermions  scattering against a static potential. Hence the assumptions underlying the boundary conformal field theory approach of Ref.s \cite{Affleck_JPA1994,AffleckProc1997}  are verified for our system.

\begin{figure}[!]
	\centering
	\includegraphics[width=.95\columnwidth]{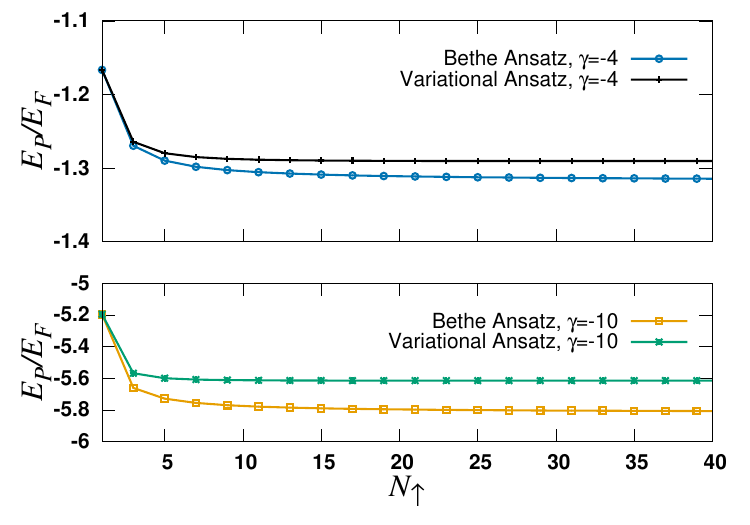}
	\caption{The polaron energy $E_P$ (i.e. the ground state energy difference $E-E_{FS}$ of the interacting and the non-interacting system) in units of the Fermi energy $E_F$  as function of the number $N_{\uparrow}$ of spin-up fermions.  Data is obtained via Bethe Ansatz and variational polaron Ansatz for $\gamma=-4$ and $\gamma=-10$.  For the variational Ansatz we observe exponential convergence, while the Bethe Ansatz converges as $1/N_\uparrow$.
	}
    \label{fig:E_N}
\end{figure}

Fig.~\ref{fig:Zfactor_N} also contains a few points obtained with DiagMC simulation. These are in excellent agreement with Bethe Ansatz. Note that the DiagMC simulation is based on a perturbative expansion around the non-interacting limit. Nonetheless, since $Z$ is finite for finite $N_{\uparrow}$, this does not pose any fundamental issue. 
 We also compare with values of $Z$ obtained with the variational Ansatz up to one particle-hole excitation. These values of $Z$ converge very rapidly to a finite value as function of $N_{\uparrow}$. So in contrast to the exact solution, the variational Ansatz predicts a non-zero quasi-particle weight in the thermodynamic limit. 
Despite the complete failure of the variational Ansatz to reproduce accurate $Z$-values at large $N_{\uparrow}$,
 it still produces excellent values for the polaron energy $E_P$ even in the thermodynamic limit.  
 The reason might lie in the rapid convergence of the variationally obtained $E_P$ and $Z$ with increasing $N_{\uparrow}$. At \emph{finite} and not too high values of $N_{\uparrow}$, \emph{both} $Z$ and $E_P$  are reasonably close to the exact answer. 
 To illustrate this, Fig.~\ref{fig:E_N} shows the polaron energy $E_P$ in units of the Fermi energy $E_F$ as function of $N_{\uparrow}$, obtained with Bethe Ansatz and variational Ansatz. 
 For $\gamma=-10$, the relative error of the varational Ansatz is about $4\%$ at large $N_{\uparrow}$. 
 The variational data for $E_P$ starts to converge around $N_{\uparrow}\simeq 5$, where the values of $Z$ are still reasonably close to the exact ones (see Fig.~\ref{fig:Zfactor_N}). In particular, we have verified that the variational result  for $E_P$ converges exponentially to its thermodynamic value as $N_\uparrow$ increases, in contrast to the exact solution, where the finite-size correction scales as $N_{\uparrow}^{-1}$.

\begin{figure}[!]
	\centering
	\includegraphics[width=.95\columnwidth]{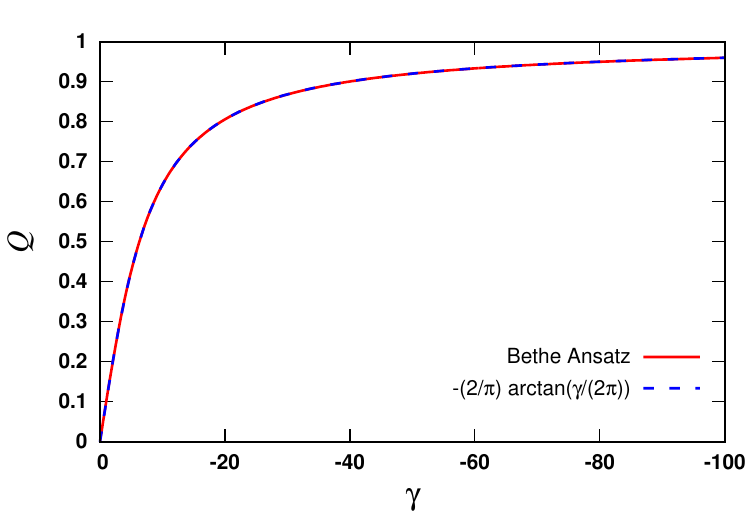}
	\caption{The charge $Q$ defined in Eq.~(\ref{eq:charge_def}) as function of the coupling $\gamma$, calculated with Bethe Ansatz (red solid line). 
	The charge changes continuously from 0 to 1, marking the continuous crossover from weak to strong coupling.  The blue dashed line corresponds to $\Delta N$, as given in Eq.~(\ref{eq:DeltaN_final}). }
    \label{fig:charge_exact}
\end{figure}

\subsection{Charge Q}\label{subsec:charge}

The charge $Q$ as function of the interaction strength $\gamma$ is shown in Fig.~\ref{fig:charge_exact}, calculated via Eqs.~(\ref{eq:charge_def}) and (\ref{eq:g2_Bethe}).  
We see that the charge $Q$ varies continuously from $Q=0$ at zero coupling to $Q=1$ in the limit of strong coupling. 
The charge is thus not quantized in one spatial dimension, in disagreement with the variational Ansatz which predicts $Q=0$.
 Importantly, we find excellent numerical agreement between the 
value of $Q$ as calculated from the correlation function $\tilde{g}_2$ and the total number $\Delta N$ of extra majority fermions surrounding the impurity   
\begin{equation}\label{eq:DeltaN_final}
    \Delta N = - \frac{2}{\pi} \arctan\left(\frac{\gamma}{2\pi}\right),\; 
\end{equation}
obtained from Eqs. (\ref{DeltaN}) and (\ref{deltaN}).

\begin{figure}[!]
	\centering
	\includegraphics[width=.95\columnwidth]{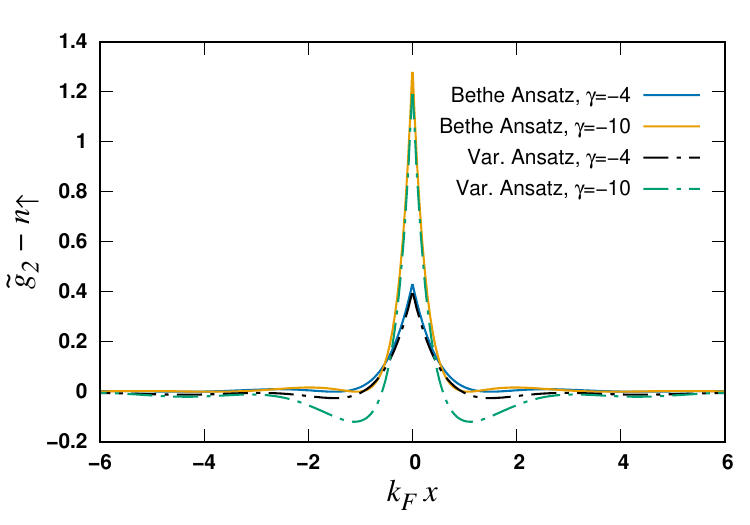}
	\caption{The pair correlation function $\tilde{g}_2(x) = g_2(x)/n_{\downarrow}$ shifted by the homogeneous spin-up density $n_{\uparrow} = k_F/\pi$ for two values of the coupling strength $\gamma$. Exact Bethe Ansatz results as well as results based on the variational Ansatz are shown. }
    \label{fig:g2}
\end{figure}

To better understand how $Q$ grows with $\gamma$ due to the deformation of the homogeneous Fermi sea because of the interaction with the spin-down impurity, we plot $\tilde{g}_2(x)-n_{\uparrow}$ in Fig.~\ref{fig:g2} for two interaction strengths. We observe that for the exact solution $\tilde{g}_2(x)-n_{\uparrow} \geq 0$ for all $x$. Integration over $x$ gives a non-zero $Q$. In the strong-coupling limit, only a Dirac-$\delta$ peak at $x=0$ survives, giving $Q=1$. The variational Ansatz (see Eq.~(\ref{eq:g2_P3})), on the other hand, gives a function $\tilde{g}_2(x)$ that oscillates around $n_{\uparrow}$ such that $Q=0$. 
The central peak around $x=0$ is reproduced remarkably well. 
For $\gamma=-10$, the variational Ansatz $\tilde{g}_2(x)$ has a pronounced dip around $k_Fx\simeq 1$. This feature is absent in the exact solution. At larger distances, the variational Ansatz predicts a significantly larger distortion of the Fermi sea compared to the exact solution. 
 
Finally, we remark that many properties of the 1D Fermi polaron are similar to those of the 3D Fermi polaron with a static impurity (i.e., impurity with infinite mass $m_{\downarrow}$). In 3D, for the mass-balanced case, $m_\downarrow = m_\uparrow$, there is a critical value of the interaction strength $(k_Fa_{\text{3D}})_c=1.11(2)$ marking a first-order transition from polaron to dimeron~\cite{polaron1}, with $a_{\text 3D}$ the 3D s-wave scattering length. Upon increasing the mass ratio $m_\downarrow / m_\uparrow$, the critical interaction strength $1/(k_Fa_{\text{3D}})_c$ shifts to higher values, so deeper into the BEC regime~\cite{Giraud_thesis}. 
At infinite mass $m_{\downarrow}$, there is no sharp transition, but rather a crossover. For this case, the 3D model is exactly solvable. 
The exact correlation function $g_2$ was calculated in Ref.~\cite{Giraud_thesis}.
Calculating the charge, we find a behaviour similar to the 1D Fermi polaron: $Q$ increases continuously from $Q=0$ in the BCS limit ($1/k_Fa_{\text{3D}}=-\infty$) to $Q=1/2$ at the unitary point ($1/k_Fa_{\text{3D}}=0$) and finally to $Q=1$ in the BEC limit ($1/k_Fa_{\text{3D}}=+\infty$).  So, just like for the 1D mass-balanced mobile Fermi polaron, the charge is not quantized. This is in agreement with a general theorem stating that the charge is not quantized for a \emph{static} impurity in a charge-compressible environment in any dimension~\cite{Chen_2018}. 
Moreover, the quasiparticle residue $Z$ of the static 3D Fermi polaron vanishes 
in the thermodynamic limit, due to the Anderson orthogonality catastrophe. The variational Ansatz, however, predicts a non-zero value of $Z$~\cite{Trefzger_2012}, similar to our results in 1D with balanced mass. \\

\section{Conclusions}

We have studied the 1D Fermi polaron problem via Bethe Ansatz, diagrammatic Monte Carlo and variational Ansatz up to one particle-hole excitation.
The quasi-particle residue $Z$ for the attractive 1D Fermi polaron decays as a power law with the number of particles $N_{\uparrow}$ of the Fermi sea, consistent with Luttinger-liquid theory. The variational Ansatz, however, predicts a finite value of $Z$ in the thermodynamic limit. We have also studied how the presence of the impurity affects the density of the spin-up sea by calculating the pair correlation function. Subtracting the homogeneous background and integrating over all distances gives the charge $Q$. The exact charge grows continuously from 0 in the non-interacting case to 1 in the strong-coupling limit. 
We also find that this $Q$ agrees with $\Delta N  = - \frac{\partial E_P}{\partial E_F}$, the number of extra majority particles attracted to the impurity, calculated from thermodynamic arguments.
The varational Ansatz, on the other hand, predicts $Q=0$ at all couplings. So, although the variational Ansatz is remarkably accurate for the energy and the effective mass, it fails even qualitatively when predicting $Q$ and $Z$  \emph{in the thermodynamic limit}. Finally, it would be interesting to study the exact pair correlation function, $Q$   and $\Delta N$ in higher dimensions and compare with the variational Ansatz.

\begin{acknowledgments}
 \noindent We are grateful to E.~Boulat, X.~Leyronas, G.~Pascual, B.~Svistunov and F.~Werner for stimulating discussions. 
 We also thank X. ~Guan for kind correspondence about the work \cite{guanPRA2016}. The authors acknowledge support from Institut Universitaire de
France, PEPR Project Dyn-1D (ANR-23-PETQ-001), ANR International
Project ANR-24-CE97-0007 QSOFT, the ANR project ANR-21-CE30-0033 LODIS and the ANR Collaborative project ANR-25-CE47-6400 LowDCertif.
This work was granted access to the HPC resources of \emph{MesoPSL} financed by the Region Ile de France and the project \emph{Equip@Meso} (reference ANR-10-EQPX-29-01) of the programme Investissements d'Avenir supervised by the Agence Nationale pour la Recherche. 
\end{acknowledgments}

\noindent \emph{Data availability.} The data presented in this article are available at \cite{data_PRAOrso}.

\bibliography{main}% Produces the bibliography via BibTeX.

\end{document}